\documentclass[seceq]{ptptex}
\usepackage{epsfig,graphicx}

\title{
Standard Model Higgs inflation: CMB, Higgs mass and quantum cosmology%
}

\author{
Andrei O. \textsc{Barvinsky}%
}

\inst{
Theory Department, Lebedev Physics Institute, Leninsky Prospect 53,
Moscow 119991, Russia\\
Department of Physics, Ludwig Maximilians University,
Theresienstrasse 37, Munich, Germany }

\abst{
We consider the inflation model generated by the Standard Model (SM)
Higgs boson having a strong non-minimal curvature coupling. This
model suggests the range of the Higgs mass $135.6\; {\rm GeV}
\lesssim M_H\lesssim 184.5\;{\rm GeV}$ entirely determined by the
lower WMAP bound on the CMB spectral index. This result is based on
the renormalization group analysis of quantum effects which make the
SM phenomenology sensitive to the current cosmological data and thus
suggest CMB measurements as a SM test complementary to the LHC
program. We show naturalness of the gradient and curvature expansion
in this model in a conventional perturbation theory range of SM. The
origin of initial conditions for inflation within the quantum
cosmology concept of the tunneling state of the Universe is also
considered. In this way a complete cosmological scenario is
obtained, which embraces the formation of initial conditions for the
inflationary background in the form of a sharp probability peak in
the distribution of the inflaton field and the ongoing generation of
the CMB spectrum on this background.}

\begin{document}

\maketitle

\section{Introduction}The goal of this  work is an attempt of
constructing a fundamental particle model accounting for an
inflationary scenario and its observable CMB spectrum. An obvious
rationale behind this is the anticipation that cosmological
observations can comprise Standard Model (SM) tests complimentary to
collider experiments -- the line of thought essentially revived by
the hope for the forthcoming discovery of the Higgs boson at LHC.
While the relatively old work \cite{we-scale} had suggested that due
to quantum effects inflation depends not only on the
inflaton-graviton sector of the system but rather is strongly
effected by the GUT contents of the particle model, the series of
papers \cite{BezShap,we,BezShap1,Wil,BezShap3} transcended this idea
to the SM ground with the Higgs field playing the role of an
inflaton. This has recovered interest in a once rather popular
\cite{non-min,we-scale,BK,KomatsuFutamase} model with the Lagrangian
of the graviton-inflaton sector
    \begin{eqnarray}
    &&{\mbox{\boldmath $L$}}(g_{\mu\nu},\varPhi)=
    \frac12\left(M_P^2+\xi|\varPhi|^2\right)R
    -\frac{1}{2}|\nabla\varPhi|^{2}
    -V(|\Phi|),                     \label{inf-grav}\\
    &&V(|\Phi|)=
    \frac{\lambda}{4}(|\varPhi|^2-v^2)^2,\,\,\,\,
    |\varPhi|^2=\varPhi^\dag\varPhi,
    \end{eqnarray}
where $\varPhi$ is a scalar field whose expectation value plays the
role of an inflaton and which has a strong non-minimal curvature
coupling with $\xi\gg 1$. Here, $M_P=m_P/\sqrt{8\pi}\approx
2.4\times 10^{18}$ GeV is a reduced Planck mass, $\lambda$ is a
quartic self-coupling of $\varPhi$, and $v$ is a symmetry breaking
scale.

The motivation for this model was  based on the observation
\cite{non-min} that the problem of an exceedingly small quartic
coupling $\lambda\sim 10^{-13}$, dictated by the amplitude of
primordial CMB perturbations \cite{CMB}, can be solved by using a
non-minimally coupled inflaton with a large value of $\xi$. Later
this model with the GUT-type sector of matter fields was used to
generate initial conditions for inflation \cite{we-scale} within the
concept of the no-boundary \cite{noboundary} and tunneling
cosmological state \cite{tunnel}. The quantum evolution with these
initial data was considered in \cite{BK,efeqmy}. There it was shown
that quantum effects are critically important for this scenario.

A similar theory was recently suggested\cite{BezShap} but with the
SM Higgs boson $\varPhi$ playing the role of an inflaton instead of
the abstract GUT setup \cite{we-scale,BK}. This work advocated the
consistency of the corresponding CMB data with WMAP observations at
the tree-level approximation of the theory, which was extended in
\cite{we} to the one-loop level. This has led to the lower bound on
the Higgs mass $M_H\gtrsim 230$ GeV, originating from the
observational restrictions on the CMB spectral index \cite{we}.
However, this conclusion which contradicts the widely accepted range
115 GeV$\leq M_H\leq 180$ GeV did not take into account $O(1)$
contributions due to renormalization group (RG) running, which
qualitatively changes the situation. This was nearly simultaneously
observed in \cite{GB08,BezShap1,Wil} where the RG improvement of the
one-loop results of \cite{we} predicted the Higgs mass range almost
coinciding with the conventional one.

Here we briefly report on the RG improvement of the one-loop results
\cite{RGHiggs} and suggest the CMB compatible range of the Higgs
mass, both boundaries of this range being determined by the lower
WMAP bound on the CMB spectral index, $n_s\gtrsim 0.94$, rather than
by perturbation theory arguments. This makes the phenomenology of
this gravitating SM essentially more sensitive to the cosmological
bounds than in \cite{BezShap1,Wil,BezShap3} and makes it testable by
the CMB observations. Then we discuss the naturalness of the
gradient and curvature expansion used for the derivation of the
above result. Finally, we consider the origin of initial conditions
for inflation within the concept of the tunneling state in quantum
cosmology. In contrast to the conventional semiclassical approach to
the minisuperspace Wheeler-DeWitt equation \cite{tunnel} this state
can be generated as a saddle-point contribution to the path integral
for the statistical sum of the microcanonical ensemble in cosmology
\cite{slih,why,qchiggs}. For massive quantum fields -- the case of
the dominant contribution of the heavy EW sector of SM -- it can be
consistently defined within the conventional UV renormalization and
RG resummation \cite{qchiggs}, and its application yields the
initial conditions for inflation in the form of a sharp probability
peak for the distribution of the initial value of the inflaton
field. In this way a complete cosmological scenario is obtained,
which embraces the formation of initial conditions for the
inflationary background and the ongoing generation of the CMB
spectrum.

\section{Quantum effects of the inflationary dynamics and CMB:
the role of the anomalous scaling}

The usual understanding of non-renormalizable theories is that
renormalization of higher-dimensional operators does not effect the
renormalizable sector of low-dimensional operators, because the
former ones are suppressed by powers of a cutoff -- the Planck mass
$M_P$ \cite{Weinberg}. Therefore, beta functions of the Standard
Model sector are not expected to be modified by gravitons. The
situation with the non-minimal coupling is more subtle. Due to
mixing of the Higgs scalar field with the longitudinal part of
gravity in the kinetic term of the Lagrangian (\ref{inf-grav}), an
obvious suppression of pure graviton loops by the effective Planck
mass, $M_P^2+\xi\varphi^2\gg M_P^2$, for large $\xi$ proliferates to
the sector of the Higgs field, so that certain parts of beta
functions are strongly damped by large $\xi$ \cite{Wil}. Therefore,
a special combination of coupling constants $\mbox{\boldmath$A$}$
which we call {\em anomalous scaling} \cite{we-scale} becomes very
small and brings down the CMB compatible Higgs mass bound. The
importance of this quantity follows from the fact observed in
\cite{we-scale,BK,we} that due to large $\xi$, quantum effects and
their CMB manifestation are universally determined by
$\mbox{\boldmath$A$}$. The nature of this quantity is as follows.

Let the model contain in addition to (\ref{inf-grav}) also a set of
scalar fields $\chi$, vector gauge bosons $A_\mu$ and spinors
$\psi$, which have an interaction with $\varPhi$ dictated by the
local gauge invariance. If we denote by $\varphi$ the inflaton --
the only nonzero component of the mean value of $\varPhi$ in the
cosmological state, then the quantum effective action of the system
takes a generic form
    \begin{equation}
    S[g_{\mu\nu},\varphi]=\int d^{4}x\,g^{1/2}
    \left(-V(\varphi)+U(\varphi)\,R(g_{\mu\nu})-
    \frac12\,G(\varphi)\,(\nabla\varphi)^2+...\right),   \label{effaction}
    \end{equation}
where $V(\varphi)$, $U(\varphi)$ and $G(\varphi)$ are the
coefficients of the derivative expansion, and we disregard the
contribution of higher-derivative operators negligible in the
slow-roll approximation of the inflation theory. In this
approximation the dominant quantum contribution to these
coefficients comes from the heavy massive sector of the model. In
particular, the masses of the physical particles and Goldstone modes
$m(\varphi)$, generated by their quartic, gauge and Yukawa couplings
with $\varphi$, give rise to the Coleman-Weinberg potential -- the
one-loop contribution to the effective potential $V$ in
(\ref{effaction}). Since $m(\varphi)\sim\varphi$, for large
$\varphi$ this potential reads
   \begin{eqnarray}
    &&V^{\rm 1-loop}(\varphi)=\sum_{
    \rm particles}
    (\pm 1)\,\frac{m^4(\varphi)}{64\pi^2}
    \,\ln\frac{m^2(\varphi)}{\mu^2}
    =\frac{\lambda\mbox{\boldmath$A$}}{128\pi^2}
    \,\varphi^4
    \ln\frac{\varphi^2}{\mu^2}+...  \label{Aviamasses}
    \end{eqnarray}
and, thus, determines the dimensionless coefficient
$\mbox{\boldmath$A$}$ -- the anomalous scaling associated with the
normalization scale $\mu$ in (\ref{Aviamasses}). Moreover, for
$\xi\gg 1$ mainly this quantity and the dominant quantum correction
to $U(\varphi)$ \cite{RGHiggs},
   \begin{eqnarray}
   U^{\rm 1-loop}(\varphi)=
    \frac{3\xi\lambda}{32\pi^2}\,\varphi^2
    \ln\frac{\varphi^2}{\mu^2}+...\, ,     \label{U1loop}
    \end{eqnarray}
determine the quantum rolling force in the effective equation of the
inflationary dynamics \cite{BK,efeqmy} and  yields the parameters of
the CMB generated during inflation \cite{we}.

Inflation and its CMB are easy to analyze in the Einstein frame of
fields $\hat g_{\mu\nu}$, $\hat\varphi$ in which the action $\hat
S[\hat g_{\mu\nu},\hat\varphi]=S[g_{\mu\nu},\varphi]$ has a minimal
coupling, canonically normalized inflaton field and the new inflaton
potential $\hat{V}=M_P^4 V(\varphi)/4U^2(\varphi)$\footnote{The
Einstein and Jordan frames are related by the equations $\hat
g_{\mu\nu}=2U(\varphi)g_{\mu\nu}/M_P^2$, $
(d\hat\varphi/d\varphi)^2=M_P^2(GU+3U'^2)/2U^2$.}, which reads at
the inflation scale as
        \begin{eqnarray}
        \hat{V}=\frac{M_P^4
    V(\varphi)}{4U^2(\varphi)}\simeq\frac{\lambda
        M_P^4}{4\,\xi^2}\,\left(1-\frac{2M_P^2}{\xi\varphi^2}+
        \frac{\mbox{\boldmath$A_I$}}{16\pi^2}
        \ln\frac{\varphi}{\mu}\right).            \label{hatVbigphi}
        \end{eqnarray}
Here the parameter $\mbox{\boldmath$A_I$}$ represents the anomalous
scaling (\ref{A0}) modified by the quantum correction to the
non-minimal curvature coupling (\ref{U1loop}),
       \begin{eqnarray}
        \mbox{\boldmath$A_I$}=\mbox{\boldmath$A$}-12\lambda=
        \frac3{8\lambda}\Big(2g^4 +
        \big(g^2 + g'^2\big)^2- 16y_t^4 \Big)-6\lambda.  \label{AI}
        \end{eqnarray}
This quantity -- which we will call {\em inflationary anomalous
scaling} -- enters the expressions for the slow-roll parameters,
$\hat\varepsilon \equiv(M_P^2/2V^2)(d\hat V/d\hat\varphi)^2$ and
$\hat\eta\equiv (M_P^2/\hat V)d^2\hat V/d\hat\varphi^2$, and
ultimately determines all the inflation characteristics. In
particular, smallness of $\hat\varepsilon$ yields the range of the
inflationary stage $\varphi>\varphi_{\rm end}$, terminating at the
value of $\hat\varepsilon$, which we chose to be
$\hat\varepsilon_{\rm end}=3/4$. Then the inflaton value at the exit
from inflation equals $\varphi_{\rm end}\simeq 2M_P/\sqrt{3\xi}$
under the natural assumption that perturbation expansion is
applicable for $\mbox{\boldmath$A_I$}/64\pi^2\ll 1$. The value of
$\varphi$ at the beginning of the inflation stage of duration $N$ in
units of the e-folding number then reads \cite{we}
    \begin{eqnarray}
    &&\varphi^2=\frac{4N}3\frac{M_P^2}{\xi}\frac{e^x-1}x, \label{xversusvarphi}\\
    &&x\equiv\frac{N
    \mbox{\boldmath$A_I$}}{48\pi^2},           \label{x}
    \end{eqnarray}
where a special parameter $x$ directly involves the anomalous
scaling $\mbox{\boldmath$A_I$}$.

This relation determines the Fourier power spectrum for the scalar
metric perturbation $\zeta$, $\Delta_{\zeta}^2(k) \equiv
<k^3\zeta_{{\bf k}}^2> = \hat V/24\pi^2M_P^4\hat\varepsilon$, where
the right-hand side is taken at the  first horizon crossing, $k=aH$,
relating the comoving perturbation wavelength $k^{-1}$ to the
e-folding number $N$,
    \begin{eqnarray}
    \Delta_{\zeta}^2(k)=
    \frac{N^2}{72\pi^2}\,\frac\lambda{\xi^2}\,
    \left(\frac{e^x-1}{x\,e^x}\right)^2.       \label{zeta}
    \end{eqnarray}
The CMB spectral index $n_s\equiv 1+d\ln\Delta_{\zeta}^2/d\ln
k=1-6\hat\varepsilon+2\hat\eta$ and the tensor to scalar ratio
$r=16\hat\varepsilon$ correspondingly read as\footnote{Note that for
$|x|\ll 1$ these predictions exactly coincide with those \cite{S83}
of the $f(R)=M_P^2(R+R^2/6M^2)/2$ inflationary model \cite{S80} with
the scalar particle (scalaron) mass $M=M_P\sqrt \lambda/\sqrt 3
\xi$.}
    \begin{eqnarray}
    &&n_s=
    1-\frac{2}{N}\, \frac{x}{e^x-1}~,           \label{ns}\\
    &&r=\frac{12}{N^2}\,
    \left(\frac{x e^x}{e^x-1}\right)^2~.          \label{r}
    \end{eqnarray}
Therefore, with the spectral index constraint $0.94 <n_s(k_0)<0.99$
(the combined WMAP+BAO+SN data at the pivot point $k_0=0.002$
Mpc$^{-1}$ corresponding to $N\simeq 60$ \cite{WMAP}) these
relations immediately give the range of anomalous scaling $-12<
\mbox{\boldmath$A_I$}<14$ \cite{we}.

On the other hand, in the Standard Model $\mbox{\boldmath$A$}$ is
expressed in terms of the masses of the heaviest particles --
$W^\pm$ boson, $Z$ boson and top quark,
    \begin{eqnarray}
    &&m_W^2=\frac14\,g^2\,\varphi^2,\;\;
    m_Z^2=\frac14\,(g^2+g'^2)\,\varphi^2,\;\;
    m_t^2=\frac12\,y_t^2\,\varphi^2,               \label{masses}
    \end{eqnarray}
and the mass of three Goldstone modes
$m_G^2=V'(\varphi)/\varphi=\lambda(\varphi^2-v^2)\simeq
\lambda\varphi^2$. Here, $g$ and $g'$ are the $SU(2)\times U(1)$
gauge couplings, $g_s$ is the $SU(3)$ strong coupling, and $y_t$ is
the Yukawa coupling for the top quark. At the inflation stage the
Goldstone mass $m_G^2$ is non-vanishing in contrast to its zero
on-shell value in the electroweak vacuum \cite{WeinbergQFT}.
Therefore, Eq. (\ref{Aviamasses}) gives the expression
   \begin{equation}
    {\mbox{\boldmath $A$}} =
    \frac3{8\lambda}\Big(2g^4 +
    \big(g^2 + g'^2\big)^2- 16y_t^4 \Big)+6\lambda.   \label{A0}
    \end{equation}
In the conventional range of the Higgs mass 115 GeV$\leq M_H\leq$
180 GeV \cite{particle} this quantity at the electroweak scale
belongs to the range $-48<\mbox{\boldmath$A$}<-20$ which strongly
contradicts the CMB range given above.

However, the RG running of coupling constants is strong enough and
drives ${\mbox{\boldmath $A$}}$  to the CMB compatible range at the
inflation scale. Below we show that the formalism of \cite{we} stays
applicable but with the electroweak ${\mbox{\boldmath $A$}}$
replaced by the running ${\mbox{\boldmath $A$}}(t)$, where
$t=\ln(\varphi/\mu)$ is the running scale of the renormalization
group (RG) improvement of the effective potential
\cite{ColemanWeinberg}.

\section{Renormalization group improvement}
According to the Coleman-Weinberg technique \cite{ColemanWeinberg}
the one-loop RG improved effective action has the form
(\ref{effaction}) with
    \begin{eqnarray}
    &&V(\varphi)=
    \frac{\lambda(t)}{4}\,Z^4(t)\,\varphi^4,  \label{RGeffpot}\\
    &&U(\varphi)=
    \frac12\Big(M_P^2
    +\xi(t)\,Z^2(t)\,\varphi^{2}\Big),      \label{RGeffPlanck}\\
    &&G(\varphi)=Z^2(t).            \label{phirenorm1}
    \end{eqnarray}
Here the running scale $t=\ln(\varphi/M_t)$ is normalized at the top
quark mass $\mu=M_t$ (we denote physical (pole) masses by capital
letters in contrast to running masses (\ref{masses}) above). The
running couplings $\lambda(t)$, $\xi(t)$ and the field
renormalization $Z(t)$ incorporate summation of powers of logarithms
and belong to the solution of the RG equations
    \begin{eqnarray}
    &&\frac{d g_i}{d t}
    =\beta_{g_i},\,\,\,\,\frac{dZ}{d t}
    =\gamma Z                   \label{renorm0}
    \end{eqnarray}
for the full set of coupling constants
$g_i=(\lambda,\xi,g,g',g_s,y_t)$ in the ``heavy'' sector of the
model with the corresponding beta functions $\beta_{g_i}$ and the
anomalous dimension $\gamma$ of the Higgs field.

An important subtlety with these $\beta$ functions is the effect of
non-minimal curvature coupling of the Higgs field. For large $\xi$
the kinetic term of the tree-level action has a strong mixing
between the graviton $h_{\mu\nu}$ and the quantum part of the Higgs
field $\sigma$ on the background $\varphi$. Symbolically it has the
structure
\[ (M_P^2+\xi^2\varphi^2)h\nabla\nabla
h+\xi\varphi\sigma\nabla\nabla h+\sigma\Box\sigma,\] which yields a
propagator whose elements are suppressed by a small $1/\xi$-factor
in all blocks of the $2\times2$ graviton-Higgs sector. For large
$\varphi\gg M_P/\sqrt\xi$, the suppression of pure graviton loops
is, of course, obvious because of the effective Planck mass squared
essentially exceeding the Einstein one, $M_P^2+\xi\varphi^2\gg
M_P^2$. Due to mixing, this suppression proliferates to the full
graviton-Higgs sector of the theory and gives the Higgs propagator
$s(\varphi)/(\Box-m_H^2)$ weighted by the suppression factor
$s(\varphi)$
    \begin{eqnarray}
    s(\varphi)=
    \frac{M_P^2+\xi\varphi^2}
    {M_P^2+(6\xi+1)\xi\varphi^2}.         \label{s}
    \end{eqnarray}

This mechanism \cite{our-ren,BK,efeqmy} modifies the beta functions
of the SM sector \cite{Wil} at high energy scales because the factor
$s(\varphi)$, which is close to one at the EW scale $v\ll M_P/\xi$,
is very small for $\varphi\gg M_P/\sqrt\xi$, $s\simeq 1/6\xi$. Such
a modification, in fact, justifies the extension beyond the scale
$M_P/\xi$ interpreted in \cite{BurgLeeTrott,BarbEsp} as a natural
validity cutoff of the theory\footnote{\label{footnote4}The
smallness of this cutoff could be interpreted as inefficiency of the
RG analysis beyond the range of trustability of the model. However,
the cutoff $M_P/\xi\ll M_P$ of \cite{BurgLeeTrott,BarbEsp} applies
to energies (momenta) of scattering processes in flat spacetime with
a small EW value of $\varphi$. For the inflation stage on the
background of a large $\varphi$ this cutoff gets modified due to the
increase in the effective Planck mass $M_P^2+\xi\varphi^2\gg M_P^2$
(and the associated decrease of the $s$-factor (\ref{s}) --
resummation of terms treated otherwise as perturbations in
\cite{BurgLeeTrott}). Thus the magnitude of the Higgs field at
inflation is not really indicative of the violation of the physical
cutoff bound (see discussion in Sects. 5 and 6 below).}.

There is an important subtlety with the modification of beta
functions which was disregarded in \cite{Wil} (and in the first
version of \cite{RGHiggs}).  Goldstone modes, in contrast to the
Higgs particle, do not have a kinetic term mixing with gravitons
\cite{BezShap3}. Therefore, their contribution is not suppressed by
the $s$-factor of the above type. Separation of Goldstone
contributions from the Higgs ones leads to the following
modification of the one-loop beta functions, which is essentially
different from that of \cite{Wil} (cf. also \cite{Clarcketal})
    \begin{eqnarray}
    &&\beta_{\lambda} = \frac{\lambda}{16\pi^2}
    \left(18s^2\lambda
    +{\mbox{\boldmath $A$}}(t)\right)
    -4\gamma\lambda,                           \label{beta-lambda}\\
    &&\beta_{\xi} =
    \frac{6\xi}{16\pi^2}(1+s^2)\lambda
    -2\gamma\xi,                 \label{beta-xi}\\
    &&\beta_{y_t} = \frac{y_t}{16\pi^2}
    \left(-\frac{2}{3}g'^2
    - 8g_s^2 +\left(1+\frac{s}2\right)y_t^2\right)
    -\gamma y_t,                                    \label{beta-y}\\
    &&\beta_{g} = -\frac{39 - s}{12}
    \frac{g^3}{16\pi^2},                     \label{beta-g}\\
    &&\beta_{g'} =
    \frac{81 + s}{12} \frac{g'^3}{16\pi^2},  \label{beta-g1}\\
    &&\beta_{g_s} =
    -\frac{7 g_s^3}{16\pi^2}.                    \label{beta-gs}
    \end{eqnarray}
Here the anomalous dimension of the Higgs field $\gamma$ is given by
a standard expression in the Landau gauge
    \begin{eqnarray}
    \gamma=\frac1{16\pi^2}\left(\,\frac{9g^2}4
    +\frac{3g'^2}4 -3y_t^2\right),                  \label{gamma}
    \end{eqnarray}
the anomalous scaling ${\mbox{\boldmath $A$}}(t)$ is defined by
(\ref{A0}) and we retained only the leading terms in $\xi\gg 1$. It
will be important in what follows that this anomalous scaling
contains the Goldstone contribution $6\lambda$, so that the full
$\beta_\lambda$ in (\ref{beta-lambda}) has a $\lambda^2$-term
unsuppressed by $s(\varphi)$ at large scale $t=\ln(\varphi/\mu)$.

The inflationary stage in units of Higgs field e-foldings is very
short, which allows one to use the approximation linear in $\Delta
t\equiv t-t_{\rm end}= \ln(\varphi/\varphi_{\rm end})$, where the
initial data point is chosen at the end of inflation $t_{\rm end}$.
Therefore, for beta functions (\ref{beta-lambda}) and
(\ref{beta-xi}) with $s\ll 1$ we have
    \begin{eqnarray}
    &&\lambda(t) = \lambda_{\rm end}\left(1
    - 4\gamma_{\rm end}\Delta t
    +\frac{\mbox{\boldmath $A$}(t_{\rm end})}{16\pi^2}\,
    \Delta t\right),                             \label{lambda-lin}\\
    &&\xi(t) = \xi_{\rm end}\Big(1
    -2\gamma_{\rm end}\Delta t
    +\frac{6\lambda_{\rm end}}{16\pi^2}\Delta t\Big),                \label{xi-lin}
    \end{eqnarray}
where $\lambda_{\rm end}$, $\gamma_{\rm end}$, $\xi_{\rm end}$ are
determined at $t_{\rm end}$ and ${\mbox{\boldmath $A$}}_{\rm
end}={\mbox{\boldmath $A$}}(t_{\rm end})$ is also a particular value
of the running anomalous scaling (\ref{A0}) at the end of inflation.

On the other hand, the RG improvement of the effective action
(\ref{RGeffpot})-(\ref{phirenorm1}) implies that this action
coincides with the tree-level action  for a new field
$\phi=Z(t)\varphi$ with running couplings as functions of
$t=\ln(\varphi/\mu)$ (the running of $Z(t)$ is slow and affects only
the multi-loop RG improvement). Then, in view of
(\ref{RGeffpot})-(\ref{RGeffPlanck}) the RG improved potential takes
at the inflation stage the form of the one-loop potential
(\ref{hatVbigphi}) for the field $\phi$ with a particular choice of
the normalization point $\mu=\phi_{\rm end}$ and all the couplings
replaced by their running values taken at $t_{\rm end}$. Therefore,
the formalism of \cite{we} can be directly applied to find the CMB
parameters of the model, which now turn out to be determined by the
running anomalous scaling ${\mbox{\boldmath $A_I$}}(t)$ taken at
$t_{\rm end}$ .

In contrast to the inflationary stage, the post-inflationary running
is very large and requires numerical simulation. We fix the $t=0$
initial conditions for the RG equations (\ref{renorm0}) at the top
quark scale $M_t =171$ GeV. For the constants $g,g'$ and
 $g_s$, they read \cite{particle}
    \begin{equation}
    g^2(0) = 0.4202,\  g'^2(0) = 0.1291,
    \ g_s^2(0) = 1.3460,                          \label{initial}
    \end{equation}
where $g^2(0)$ and $g'^2(0)$ are obtained by a simple one-loop RG
flow from the conventional values of $\alpha(M_Z)\equiv
g^2/4\pi=0.0338$, $\alpha'(M_Z)\equiv g'^2/4\pi=0.0102$ at
$M_Z$-scale, and the value $g_s^2(0)$ at $M_t$ is generated by the
numerical program of \cite{QCDfromZtotop}. The analytical algorithm
of transition between different scales for $g_s^2$ was presented in
\cite{DV}. For the Higgs self-interaction constant $\lambda$ and for
the Yukawa top quark interaction constant $y_t$ the initial
conditions are determined by the pole mass matching scheme
originally developed in \cite{top} and presented in the Appendix of
\cite{espinosa}.

The initial condition $\xi(0)$ follows from the CMB normalization
(\ref{zeta}), $\Delta_{\zeta}^2\simeq 2.5\times 10^{-9}$ at the
pivot point $k_0=0.002$ Mpc$^{-1}$, which we choose to correspond to
$N\simeq 60$.\cite{WMAP} This yields the following estimate on the
ratio of coupling constants
    \begin{equation}
\frac{1}{Z_{\rm in}^2}\frac{\lambda_{\rm in}}{\xi^2_{\rm in}}
    \simeq 0.5\times 10^{-9}
    \left(\frac{x_{\rm in}\,\exp x_{\rm in}}
    {\exp x_{\rm in}-1}\right)^2              \label{final}
    \end{equation}
at the moment of the first horizon crossing for $N=60$ which we call
the ``beginning'' of inflation and label by $t_{\rm
in}=\ln(\varphi_{\rm in}/M_t)$ with $\varphi_{\rm in}$ defined by
(\ref{xversusvarphi}). Thus, the RG equations (\ref{renorm0}) for
six couplings $(g,g',g_s,y_t,\lambda,\xi)$ with five initial
conditions and the final condition for $\xi$ uniquely determine the
needed RG flow.

The RG flow covers also the inflationary stage from the
chronological end of inflation $t_{\rm end}$  to $t_{\rm in}$. At
the end of inflation we choose the value of the slow roll parameter
$\hat\varepsilon=3/4$, and $\varphi_{\rm end}\equiv M_t e^{t_{\rm
end}}\simeq M_P\sqrt{4/3\xi_{\rm end}}$. Thus, the duration of
inflation in units of inflaton field e-foldings $t_{\rm in}-t_{\rm
end}=\ln(\varphi_{\rm in}/\varphi_{\rm end})\simeq\ln N/2\sim 2$ is
very short relative to the post-inflationary evolution $t_{\rm
end}\sim 35$. The approximation linear in logs implies the bound
$|{\mbox{\boldmath $A_I$}}(t_{\rm end})|\Delta t/16\pi^2\ll 1$,
which in view of $\Delta t<t_{\rm in}-t_{\rm end}\simeq \ln N/2$
holds for $|{\mbox{\boldmath $A_I$}}(t_{\rm end})|/16\pi^2\ll 0.5$.

\section{CMB compatible bounds on the Higgs mass}
The running of ${\mbox{\boldmath $A$}}(t)$ strongly depends on the
behavior of $\lambda(t)$. For small Higgs masses the usual RG flow
in SM leads to an instability of the EW vacuum caused by negative
values of $\lambda(t)$ in a certain range of $t$
\cite{Sher,espinosa}. The same happens in the presence of
non-minimal curvature coupling.
\begin{figure}[h]
\centerline{\epsfxsize 12cm \epsfbox{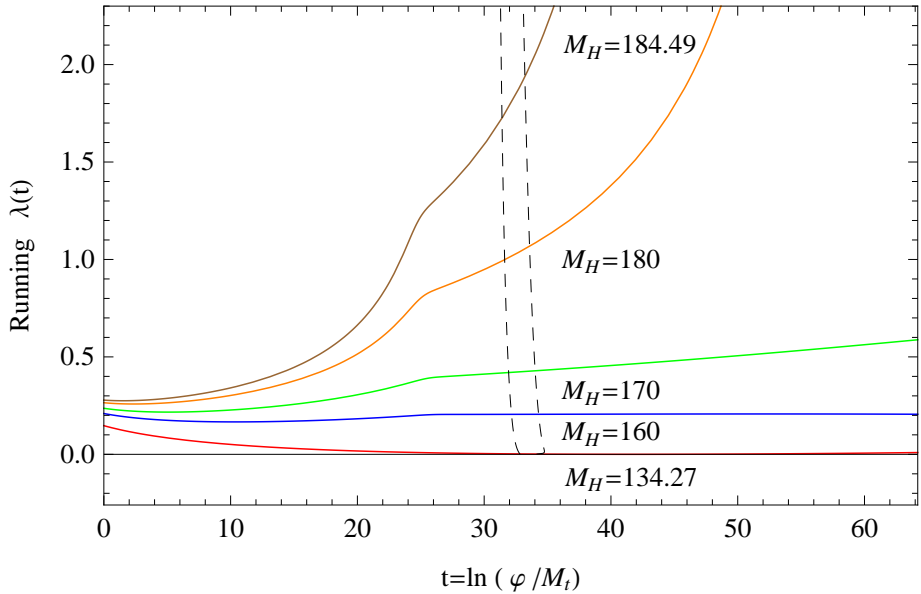}}
\caption{\small Running $\lambda(t)$ for five values of the Higgs
mass above the instability threshold. Dashed curves mark the
boundaries of the inflation domain $t_{\rm end}\leq t\leq t_{\rm
in}$.
 \label{Fig.1}}
\end{figure}
The numerical solution for $\lambda(t)$ is shown in Fig.1 for five
values of the Higgs mass and the value of top quark mass $M_t=171$
GeV. The lowest one corresponds to the boundary of the instability
window,
    \begin{equation}
    M_H^{\rm inst}\simeq 134.27\; {\rm GeV},      \label{criticalmass}
    \end{equation}
for which $\lambda(t)$ bounces back to positive values after
vanishing at $t_{\rm inst}\sim 41.6$ or $\varphi_{\rm inst}\sim 80
M_P$. It turns out that the corresponding $\xi(t)$ is nearly
constant and is about $5000$ (see below), so that the factor
(\ref{s}) at $t_{\rm inst}$ is very small $s\simeq 1/6\xi\sim
0.00005$. Thus the situation is different from the usual Standard
Model with $s=1$, and numerically the critical value turns out to be
higher than the known SM stability bound $\sim 125$ GeV
\cite{espinosa}.

Fig.1 shows that near the instability threshold $M_H=M_H^{\rm inst}$
the running coupling $\lambda(t)$ stays very small for all scales
$t$ relevant to the observable CMB. This follows from the fact that
the positive running of $\lambda(t)$ caused by the term $(18
s^2+6)\lambda^2$ in $\beta_\lambda$, (\ref{beta-lambda}), is much
slower for $s\ll 1$ than that of the usual SM driven by the term
$24\lambda^2$.

\begin{figure}[h]
\centerline{\epsfxsize 12cm \epsfbox{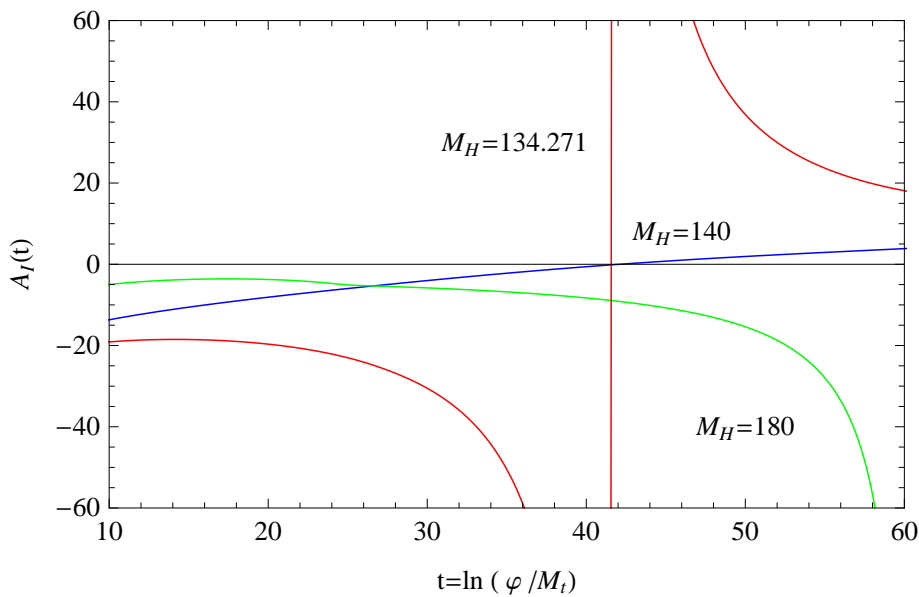}} \caption{\small
Running anomalous scaling for the critical Higgs mass (the red curve
with a vertical segment at the singularity with $t_{\rm inst}\sim
41.6$) and for two masses in the stability domain (blue and green
curves).
 \label{Fig.2}}
\end{figure}

\begin{figure}[h]
\centerline{\epsfxsize 10cm \epsfbox{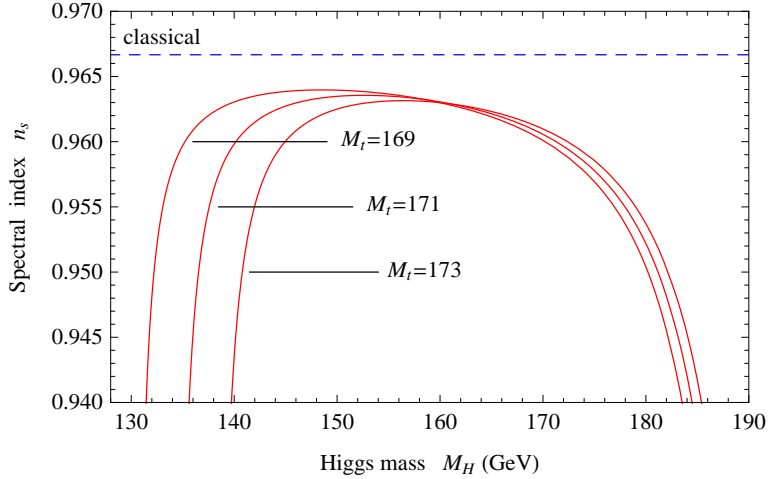}}
\caption{\small The spectral index $n_s$ as a function of the Higgs
mass $M_H$ for three values of the top quark mass.
 \label{Fig.3}}
\end{figure}

For all Higgs masses in the range $M_H^{\rm inst}=134.27$ GeV
$<M_H<185$ GeV the inflation range $t_{\rm end}<t<t_{\rm in}$ is
always below $t_{\rm inst}=41.6$ , so that from Fig.~2
${\mbox{\boldmath $A_I$}}(t)$ is always negative during inflation.
Its running explains the main difference from the results of
one-loop calculations \cite{we}. ${\mbox{\boldmath $A_I$}}(t)$ runs
from big negative values ${\mbox{\boldmath $A_I$}}(0)<-20$ at the
electroweak scale to small also negative values at the inflation
scale below $t_{\rm inst}$. This makes the CMB data compatible with
the generally accepted Higgs mass range. Indeed, the knowledge of
the RG flow immediately allows one to obtain ${\mbox{\boldmath
$A_I$}}(t_{\rm end})$ and $x_{\rm end}$ and thus find the parameters
of the CMB power spectrum (\ref{ns})-(\ref{r}) as functions of
$M_H$. The parameter of primary interest -- spectral index -- is
given by Eq. (\ref{ns}) with $x=x_{\rm end}\equiv N{\mbox{\boldmath
$A_I$}}(t_{\rm end})/48\pi^2$ and depicted in Fig.\ref {Fig.3}. Even
for low values of Higgs mass above the stability bound, $n_s$ falls
into the range admissible by the CMB constraint existing now at the
$2\sigma$ confidence level (based on the combined WMAP+BAO+SN data
\cite{WMAP}) $0.94 <n_s(k_0)<0.99$.

The spectral index drops below 0.94 only for large $x_{\rm end}<0$
or large negative ${\mbox{\boldmath $A_I$}}(t_{\rm end})$, which
happens only when $M_H$ either approaches the instability bound or
exceeds 180 GeV at the decreasing branch of the $n_s$ graph. Thus,
we get the lower and upper bounds on the Higgs mass, which both
follow from the lower bound of the CMB data. Numerical analysis for
the corresponding $x_{\rm end}\simeq -1.4$ gives for $M_t=171$ GeV
the range of CMB compatible Higgs mass
    \begin{equation}
    135.62\; {\rm GeV}\lesssim M_H
    \lesssim 184.49\; {\rm GeV}.       \label{CMBmass}
    \end{equation}
Both bounds belong to the nonlinear domain of the equation
(\ref{ns}) with $x_{\rm end}\simeq-1.4$, but the quantity
$|\mbox{\boldmath $A_I$}(t_{\rm end})|/16\pi^2=0.07\ll 0.5$
satisfies the restriction mentioned above, and their calculation is
still in the domain of our linear in logs approximation.

The upper bound on $n_s$ does not generate restrictions on $M_H$.
The lower CMB bound in (\ref{CMBmass}) is slightly higher than the
instability bound $M_H^{\rm inst}=134.27$ GeV. In turn, this bound
depends on the initial data for weak and strong couplings and on the
top quark mass $M_t$ which is known with less precision. The above
bounds were obtained for $M_t=171$ GeV. Results for the neighboring
values $M_t=171\pm2$ GeV are presented in Fig. \ref{Fig.3} to show
the pattern of their dependence on $M_t$.

Finally let us focus on the running of $\xi(t)$. It is very slow for
low values of the Higgs mass near the instability threshold. Of
course, this follows from the smallness of the running $\lambda(t)$
in this domain. Another property of the $\xi$-behavior is that the
normalization of the power spectrum leads to a value $\xi\sim 5000$
for small Higgs masses, which is smaller than the old
estimate\cite{non-min,we-scale,BK,efeqmy,BezShap,we} $\sim 10^4$ .
This is caused by a decrease of $\lambda(t)$ which at $t_{\rm in}$
becomes much smaller than $\lambda(0)$ \cite{Wil}.

\section{Gradient and curvature expansion cutoff and naturalness}
The expression (\ref{effaction}) is a truncation of the curvature
and derivative expansion of the full effective action. It was
repeatedly claimed that with large $\xi$ the weak field version of
this expansion on flat (and empty) space background has a cutoff
$4\pi M_P/\xi$ \cite{BurgLeeTrott,BarbEsp}. This scale is
essentially lower than the Higgs field during inflation $\varphi\sim
M_P/\sqrt\xi$ and, therefore, seems to invalidate predictions based
on (\ref{effaction}) without unnatural suppression of
higher-dimensional operators\cite{GK}. The attempt to improve the
situation by transition to the Einstein frame
\cite{LernerMcDonald,BMSS} was claimed to fail
\cite{BurgLeeTrott1,Hertzberg} in view of a multiplet nature of the
Higgs field involving Nambu-Goldstone modes.

Here we show that these objections against naturalness are not quite
conclusive. First, as mentioned above, a big value of $\varphi$
during inflation is not really indicative of a large physical scale
of the problem. In contrast to curvature and energy density the
inflaton itself is not a physical observable, but rather a
configuration space coordinate of the model. Secondly, we show now
that the inflation scale actually lies below the gradient expansion
cutoff, and this justifies naturalness of the obtained results. No
transition to another conformal frame is needed for that, but rather
the resummation accounting for transition to large $\varphi$
background.

Indeed, the main peculiarity of the model (\ref{inf-grav}) is that
in the background field method with small derivatives the role of
the effective Planck mass is played by $(M_P^2+\xi\varphi^2)^{1/2}$.
The power-counting method \cite{BurgLeeTrott} underlying the
derivation of the cutoff $4\pi M_P/\xi$ also applies here but with
the Planck mass $M_P$ replaced by the effective one,
$M_P\to(M_P^2+\xi\varphi^2)^{1/2}>\sqrt\xi\varphi$. The resulting
cutoff is thus bounded from below by
    \begin{equation}
    \Lambda(\varphi)=\frac{4\pi\varphi}{\sqrt\xi},  \label{cutoff}
    \end{equation}
and this bound can be used as a {\em running} cutoff of the gradient
and curvature expansion. The origin of this cutoff can be
demonstrated in the one-loop approximation. When calculated in the
{\em Jordan} frame, the one-loop divergences quadratic in the
curvature $R$ have a strongest in $\xi$ contribution (this can be
easily deduced from the Appendix of \cite{RGHiggs})
    \begin{equation}
    \xi^2\frac{R^2}{16\pi^2}.     \label{counterterm}
    \end{equation}
As compared to the tree-level part linear in the curvature $\sim
(M_P^2+\xi\varphi^2)R$, the one loop $R^2$-term turns out to be
suppressed by the above cutoff factor
$16\pi^2(M_P^2+\xi\varphi^2)/\xi^2\simeq\Lambda^2$.

The on-shell curvature estimate at the inflation stage reads $R\sim
V/U\sim\lambda\varphi^2/\xi$, so that the resulting curvature
expansion runs in powers of
    \begin{equation}
    \frac{R}{\Lambda^2}
    \sim\frac\lambda{16\pi^2}      \label{curvatureexpansion}
    \end{equation}
and remains efficient in the usual perturbation theory range of SM,
$\lambda/16\pi^2\ll 1$. This works perfectly well in our Higgs
inflation model, because in the full CMB compatible range of the
Higgs mass $\lambda<2$ (see Fig.\ref{Fig.1}).

From the viewpoint of the gradient expansion for $\varphi$ this
cutoff is even more efficient. Indeed, the inflaton field gradient
can be expressed in terms of the inflaton potential $\hat V$ and the
inflation smallness parameter $\hat\varepsilon$ taken in the
Einstein frame,
$\dot\varphi\simeq(\varphi^2/M_P^2)(\xi\hat\varepsilon\hat{V}/18)^{1/2}$.
With $\hat V\simeq\lambda M_P^4/4\xi^2$ this immediately yields the
gradient expansion in powers of
    \begin{equation}
    \frac{\partial}{\Lambda}
    \sim\frac1\Lambda\frac{\dot\varphi}\varphi\simeq
    \frac{\sqrt\lambda}{48\pi}
    \sqrt{2\hat\varepsilon},                \label{partialexpansion}
    \end{equation}
which is even better than (\ref{curvatureexpansion}) by the factor
ranging from $1/N$ at the beginning of inflation to $O(1)$ at the
end of it.

Eqs. (\ref{curvatureexpansion}) and (\ref{partialexpansion}) justify
the effective action truncation in (\ref{effaction}) in the
inflationary domain. Thus only multi-loop corrections to the
coefficient functions $V(\varphi)$, $U(\varphi)$ and $G(\varphi)$
might stay beyond control in the form of higher-dimensional
operators $(\varphi/\Lambda)^n$ and violate the flatness of the
effective potential necessary for inflation. However, in view of the
form of the running cutoff (\ref{cutoff}) they might be large, but
do not affect the shape of these coefficient functions because of
the field independence of the ratio $\varphi/\Lambda$. Only the
logarithmic running of couplings in
(\ref{RGeffpot})-(\ref{phirenorm1}) controlled by RG dominates the
quantum input in the inflationary dynamics and its CMB
spectra\footnote{Like the logarithmic term of (\ref{hatVbigphi})
which dominates over the nearly flat classical part of the inflaton
potential and qualitatively modifies tree-level predictions of the
theory \cite{we}.}.

In this connection the results of the recent work \cite{BMSS} might
be helpful. It claims that due to asymptotic shift symmetry of the
model (or asymptotic scale invariance in the Jordan frame) the
field-dependent cutoff at the inflation scale is much higher than
(\ref{cutoff}) and is given by $\Lambda_E\sim \sqrt\xi\varphi$,
which strongly supports naturalness of Higgs inflation along with
its consistency at the reheating and Big Bang stages. The difference
from (\ref{cutoff}) can be explained by the fact that, in contrast
to our Jordan frame calculations, the quantum corrections in
\cite{BMSS} were analyzed in the Einstein frame. In this frame, in
particular, the strongest curvature squared counterterm is
$O(1)R^2/16\pi^2$ rather than (\ref{counterterm}) (see
\cite{our-ren}). Of course, modulo the conformal anomaly
contribution, which only effects the log arguments and cannot be
responsible for the onshell discrepancy in these counterterms, the
physical results should be equivalent in both parameterizations
\cite{RGHiggs}. This can be an indication of intrinsic cancelations
which are not manifest in the Jordan frame and which could
effectively raise the cutoff from its naive value (\ref{cutoff}) to
$\Lambda_E$. This would also justify suppression of higher
dimensional operators in the coefficient functions $V,U,G$ mentioned
above. However, verification of frame equivalence of the physical
results should be based on gauge and parametrization invariant
definition of CMB observables which is currently under study.

\section{Quantum cosmology origin of Higgs inflation}
Now we want to show that, in addition to the good agreement of the
CMB spectrum with the observational data, this model can also
describe the mechanism of generating the cosmological {\em
background} itself upon which these perturbations exist. This
mechanism consists in the formation of the initial conditions for
inflation in the form of a sharp probability peak in the
distribution function of the initial value of the inflaton field.
Such a distribution naturally arises in quantum cosmology within two
known suggestions for the wavefunction of the Universe -- the
no-boundary Hartle-Hawking prescription \cite{noboundary} and the
prescription of the tunneling cosmological wavefunction
\cite{tunnel}. Though originally these prescriptions did not have a
good justification at the level of a consistent operator
quantization in the physical spacetime with the Lorentzian
signature\footnote{The no-boundary state was put forward as a formal
Euclidean quantum gravity path integral \cite{noboundary}, whereas
the tunneling prescription existed merely as a semiclassical
solution of the minisuperspace Wheeler-DeWitt equation
\cite{tunnel}.}, recently they were both derived as saddle-point
contributions to the statistical sum of the microcanonical ensemble
in cosmology (described by the density matrix which is just the
projector onto the space of solutions of the Wheeler-DeWitt
equations)\cite{why,slih,qchiggs}.

As it was recently shown \cite{qchiggs,Quarks10}, specifically the
tunneling state gets realized for heavy massive quantum
fields\footnote{In contrast to the case of massless conformal fields
which give rise to the thermal version of the no-boundary
state.\cite{Quarks10}} -- exactly the case of the Higgs inflation
with quantum corrections generated by a heavy EW sector of the
Standard Model. Within the relevant inverse mass expansion (which is
just the slow-roll approximation in cosmology or the expansion in
terms of higher-dimensional operators of the effective field theory)
the distribution function of the inflaton field turns out to be
given by the exponentiated {\em Euclidean effective} action,
$\rho_{\rm tunnel}(\varphi)=\exp\big(+S_E(\varphi)\big)$
\footnote{Note the plus sign in the exponential, which is different
from the minus sign corresponding to the no-boundary
case\cite{qchiggs,Quarks10}.}. This action -- the Euclidean version
of (\ref{effaction}) -- is supposed to be calculated on the
(quasi)-de Sitter instanton of a spherical topology $S^4$ of the
size determined by the magnitude of the inflaton field and its
effective inflaton potential. For our model of Higgs inflation with
a non-minmal curvature coupling this is the Einstein frame potential
(\ref{hatVbigphi}), and the tunneling distribution function
    \begin{eqnarray}
    &&\rho_{\rm tunnel}(\varphi)=
    \exp\left(-\frac{24\pi^2M_{\rm P}^4}
    {\hat V(\varphi)}\right)                \label{partitionphi}
    \end{eqnarray}
describes the ensemble of (quasi)de Sitter universes starting to
expand with various values of the effective cosmological constant
$\hat\Lambda=\hat V(\varphi)/M_P^2$. So the question is whether this
distribution can have a sharp probability peak at some appropriate
value of the inflaton field $\varphi_0$ with which the Universe as a
whole starts its evolution. The shape and the magnitude of $\hat V$
depicted in Fig.4 for several values of the Higgs mass clearly
indicates the existence of such a peak.

\begin{figure}[h]
\centerline{\epsfxsize 12cm \epsfbox{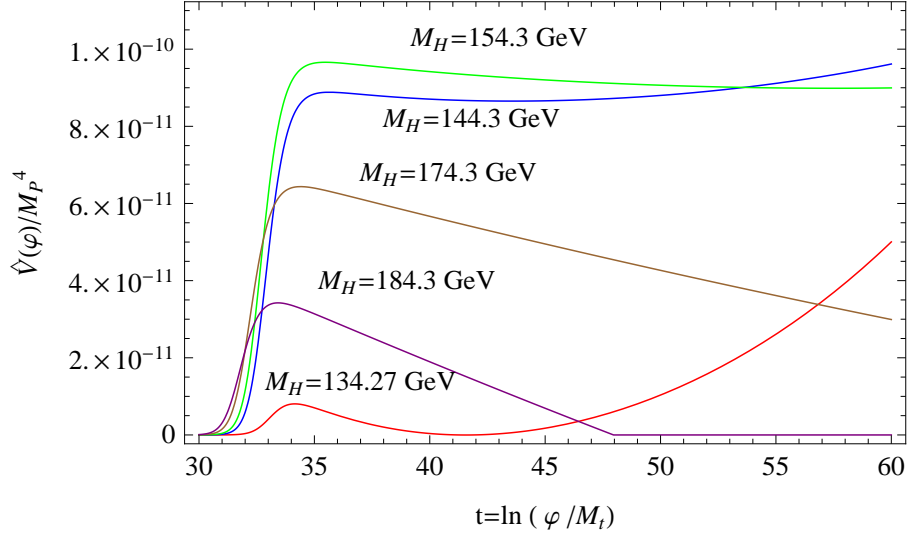}}
\caption{\small The succession of effective potential graphs above
the EW vacuum instability threshold $M_{\rm H}^{\rm inst}=134.27\
{\rm GeV}$ up to $M_{\rm H}=184.3$ GeV showing the occurrence of a
metastable vacuum followed for high $M_{\rm H}$ by the formation of
a negative slope branch. Local peaks of $\hat V$ situated at
$t=34\div35$ grow with $M_{\rm H}$ for $M_{\rm H}\lesssim 160$ GeV
and start decreasing for larger $M_{\rm
  H}$.
 \label{Fig.4}}
\end{figure}

Indeed, the negative of the inverse potential damps to zero after
exponentiation the probability of those values of $\varphi$ at which
$\hat V(\varphi)=0$ and, vice versa, enhances the probability at the
positive maxima of the potential. The pattern of this behavior with
growing Higgs mass $M_{\rm H}$ is as follows.

\begin{figure}[h]
\centerline{\epsfxsize 12cm \epsfbox{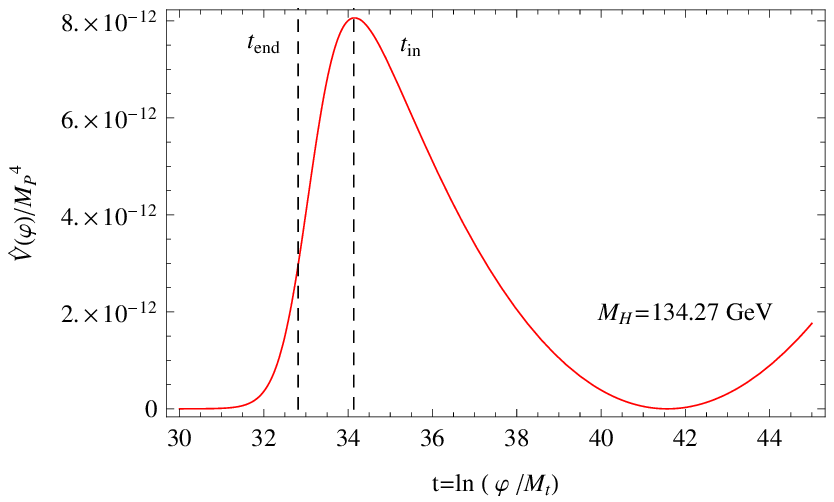}}
\caption{\small The effective potential for the instability
threshold $M_{\rm H}^{\rm inst}=134.27$ GeV. A false vacuum occurs
at the instability scale $t_{\rm inst}\simeq 41.6$, $\varphi\sim 80
M_{\rm P}$. An inflationary domain for a $N=60$ CMB perturbation
(ruled out by the WMAP bounds) is marked by dashed lines.
\label{Fig.5}}
\end{figure}

We begin with the EW vacuum instability threshold
\cite{espinosa,Sher} which exists in this gravitating SM at $M_{\rm
H}^{\rm inst}\approx 134.27$ GeV \cite{RGHiggs} and which is
slightly lower than the CMB compatible range of the Higgs mass
($M_{\rm H}^{\rm inst}$ is chosen in Fig. \ref{Fig.5} and for the
lowest curve in Fig. \ref{Fig.4}). The potential $\hat V(\varphi)$
drops to zero at $t_{\rm inst}\simeq 41.6$, $\varphi_{\rm inst}\sim
80 M_{\rm P}$, and forms a false vacuum \cite{RGHiggs} separated
from the EW vacuum by a large peak at $t\simeq 34$. Correspondingly,
the probability of creation of the Universe with the initial value
of the inflaton field at the EW scale $\varphi=v$ and at the
instability scale $\varphi_{\rm inst}$ is damped to zero, while the
most probable value belongs to this peak. The inflationary stage of
the formation of the pivotal $N=60$ CMB perturbation (from the
moment $t_{\rm in}$ of the first horizon crossing until the end of
inflation $t_{\rm end}$), which is marked by dashed lines in
Fig.\ref{Fig.5}, lies to the left of this peak. This conforms to the
requirement of the chronological succession of the initial
conditions for inflation and the formation of the CMB spectra.

The above case is, however, below the CMB-compatible range of
$M_{\rm H}$ and was presented here only for illustrative purposes.
An important situation occurs at higher Higgs masses from the lower
CMB bound on $M_{\rm H}\simeq 135.6$ GeV until about 160 GeV. Here
we get a family of a metastable vacua with $\hat V>0$. An example is
the plot with $M_{\rm H}=144.3$ GeV in Fig. \ref{Fig.4}. Despite the
shallowness of this vacuum the small maximum of $\hat V$ at $t\simeq
35$ generates via (\ref{partitionphi}) a sharp probability peak for
the initial inflaton field. This follows from an extremely small
value of $\hat V/M_{\rm P}^4\sim 10^{-11}$, the reciprocal of which
generates a rapidly changing exponential of (\ref{partitionphi}).

For even larger $M_{\rm H}$ these metastable vacua get replaced by a
negative slope of the potential which interminably decreases to zero
at large $t$ (at least within the perturbation theory range of the
model), see Fig.~\ref{Fig.4}. Therefore, for large $M_{\rm H}$ close
to the upper CMB bound 185 GeV, the probability peak of
(\ref{partitionphi}) gets separated from the non-perturbative domain
of large over-Planckian scales due to a fast drop of $\hat
V\sim\lambda/\xi^2$ to zero. This, in turn, follows from the fact
that $\xi(t)$ grows much faster than $\lambda(t)$ when they both
start approaching their Landau pole \cite{RGHiggs}.

The location $\varphi_0$ of the probability peak and its quantum
width can be found in analytical form, and their derivation shows
the crucial role of the running $\mbox{\boldmath$A_I$}(t)$ for the
formation of initial conditions for inflation. Indeed, the
exponential of (\ref{partitionphi}) for $M_{\rm P}^2/\xi\varphi^2\ll
1$ in view of the RG equations for $\lambda$ and $\xi$ with beta
functions (\ref{beta-lambda})--(\ref{beta-xi}) has an extremum
satisfying the equation
    \begin{eqnarray}
    \varphi\frac{d\varGamma}{d\varphi}=\frac{d\varGamma}{dt}
    =-\frac{6\xi^2}
    \lambda\left(\mbox{\boldmath$A_I$}
    +\frac{64\pi^2M_{\rm P}^2}{\xi Z^2\varphi^2}\right)=0,
    \end{eqnarray}
where we neglect higher order terms in $M_{\rm P}^2/\xi
Z^2\varphi^2$ and $\mbox{\boldmath$A_I$}/64\pi^2$ (extending beyond
the one-loop order). Here, $\mbox{\boldmath$A_I$}$ is the anomalous
scaling  (\ref{Aviamasses}) and (\ref{AI}) which is negative in the
domain of interest and, thus, guarantees the existence of the
solution for the probability peak,
    \begin{eqnarray}
    \varphi^2_0=
    \left.-\frac{64\pi^2M_{\rm P}^2}{\xi \mbox{\boldmath$A_I$}Z^2}
    \,\right|_{\;t=t_0}.                               \label{root}
    \end{eqnarray}
In the CMB-compatible range of $M_{\rm H}$ its running starts from
the range $-36\lesssim\mbox{\boldmath$A_I$}(0)\lesssim-23$ at the EW
scale and reaches small but still negative values in the range
$-11\lesssim\mbox{\boldmath$A_I$}(t_{\rm end})\lesssim -2$ at the
inflation scale. Also, the running of $\mbox{\boldmath$A_I$}(t)$ and
$Z(t)$ is very slow -- the quantities belonging to the two-loop
order -- and the duration of inflation is very short $t_0\sim t_{\rm
in}\simeq t_{\rm end}+2$. Therefore, $\mbox{\boldmath$A_I$}(t_{\rm
0})\simeq\mbox{\boldmath$A_I$}(t_{\rm end})$, and these estimates
apply also to $\mbox{\boldmath$A_I$}(t_{\rm 0})$. As a result, the
second derivative of the tunneling on-shell action is positive and
very large, $d^2\varGamma_-/dt^2\simeq
    -(12\xi^2/\lambda) \mbox{\boldmath$A_I$}\gg 1$,
which gives an extremely small value of the quantum width of the
probability peak,
    \begin{eqnarray}
    \frac{\Delta\varphi^2}{\varphi^2_0}
    =-\left.\frac\lambda{12\xi^2}
    \frac1{\mbox{\boldmath$A_I$}}\right|_{\,t=t_0}\sim
    10^{-10}.
    \end{eqnarray}
This width is about $(24\pi^2/|{\mbox{\boldmath$A_I$}}|)^{1/2}$
times -- one order of magnitude -- higher than the CMB perturbation
at the pivotal wavelength $k^{-1}=500$ Mpc (which we choose to
correspond to $N=60$). The point $\varphi_{\rm in}$ of the horizon
crossing of this perturbation (and other CMB waves with different
$N$'s) easily follows from equation (\ref{xversusvarphi}) which in
view of $\mbox{\boldmath$A_I$}(t_{\rm
0})\simeq\mbox{\boldmath$A_I$}(t_{\rm end})$ takes the form
    \begin{eqnarray}
    &&\frac{\varphi^2_{\rm in}}{\varphi_0^2}=
    1-\exp\left(-N\frac{|\mbox{\boldmath$A_I$}
    (t_{\rm end})|}{48\pi^2}\right).            \label{phi0phiin}
    \end{eqnarray}
It indicates that for wavelengths longer than the pivotal one the
instant of horizon crossing approaches the moment of ``creation'' of
the Universe, but always stays chronologically succeeding it, as it
must.

\section{Conclusions and discussion}
We have found that the considered model looks remarkably consistent
with CMB observations in the Higgs mass range
    \begin{equation}
    135.6\; {\rm GeV} \lesssim M_H
    \lesssim 184.5\;{\rm GeV},                \label{range}
    \end{equation}
which is very close to the widely accepted range dictated by
electroweak vacuum stability and perturbation theory bounds.

This result is based on the observation that for large $\xi\gg 1$
the effect of SM phenomenology on inflation is universally encoded
in one quantity -- the anomalous scaling ${\mbox{\boldmath $A_I$}}$.
It was earlier suggested in \cite{we-scale} for a generic gauge
theory, and in SM it is dominated by contributions of heavy
particles -- ($W^\pm$, $Z$)-bosons, top quark and Goldstone modes.
This quantity is forced to run in view of RG resummation of leading
logarithms, and this running raises a large negative EW value of
${\mbox{\boldmath $A_I$}}$ to a {\em small negative} value at the
inflation scale. Ultimately this leads to the admissible range of
Higgs masses  very close to the conventional SM range.

Qualitatively these conclusions are close to those of \cite{Wil} and
\cite{BezShap3}, though our bounds on the SM Higgs mass are much
more sensitive to the CMB data. Therefore, the latter can be
considered as a test of the SM theory complimentary to LHC and other
collider experiments. The source of this difference from
\cite{Wil,BezShap3} can be ascribed to the gauge and parametrization
(conformal frame) dependence of the off-shell effective action
(along with the omission of Goldstone modes contribution in
\cite{Wil}) -- an issue which is discussed in much detail in
\cite{RGHiggs} and which is expected to be resolved in future
publications.

We have also shown the naturalness of the gradient and curvature
expansion in this model, which is guaranteed within the conventional
perturbation theory range of SM, $\lambda/16\pi^2\ll 1$, and holds
in the whole range of the CMB compatible Higgs mass (\ref{range}).
This result is achieved by the background field resummation of weak
field perturbation theory leading to the replacement of the
fundamental Planck mass in the known cutoff $4\pi M_P/\xi$
\cite{BurgLeeTrott,BarbEsp} by the effective one. Partly (modulo
corrections to inflaton potential, which are unlikely to spoil its
shape) this refutes objections of \cite{BurgLeeTrott,BarbEsp} based
on the analysis of scattering amplitudes in EW vacuum background.
Smallness of the cutoff in this background does not contradict
physical bounds on the Higgs mass originating from CMB data for the
following reasons. Determination of $M_H$ of course takes place at
the TeV scale much below the non-minimal Higgs cutoff $4\pi
M_P/\xi$, whereas inflationary dynamics and CMB formation occur for
$\lambda/16\pi^2\ll 1$ below the {\em running} cutoff
$\Lambda(\varphi)=4\pi\varphi/\sqrt\xi$. It is the phenomenon of
inflation which due to exponentially large stretching brings these
two scales in touch and allows us to probe the physics of underlying
SM by CMB observations at the 500 Mpc wavelength scale.

We also applied the quantum cosmology concept to derive initial
conditions in this particular model of inflation. Specifically, we
used the recently suggested path integral formulation of the
tunneling cosmological state \cite{qchiggs} , which admits a
consistent renormalization scheme and becomes indispensable in the
case when quantum effects play a dominant role. In this way a
complete cosmological scenario was obtained, embracing the formation
of initial conditions for the inflationary background (in the form
of a sharp probability peak in the inflaton field distribution) and
the ongoing generation of the CMB perturbations on this background.
Interestingly, the behavior of the running anomalous scaling
$\mbox{\boldmath$A_I$}(t)<0$, which is crucially important for the
CMB formation and the corresponding Higgs mass bounds, also
guarantees the existence of the obtained probability peak
\cite{qchiggs}. The quantum width of this peak is one order of
magnitude higher than the amplitude of the CMB spectrum at the
pivotal wavelength, which could entail interesting observational
consequences. Unfortunately, this quantum width is hardly measurable
directly because it corresponds to an infinite wavelength
perturbation (a formal limit of $N\to\infty$ in (\ref{phi0phiin})),
but indirect effects of this quantum trembling of the cosmological
background deserve further study.

To summarize, the obtained results bring to life a convincing
unification of quantum cosmology with the particle phenomenology of
the SM, inflation theory, and CMB observations. They support the
hypothesis that an appropriately extended Standard Model \cite{dark}
can be a consistent quantum field theory all the way up to quantum
gravity and perhaps explain the fundamentals of all major phenomena
in early and late cosmology.

\section*{Acknowledgements}
This paper is based on a series of works done together with
A.Yu.Kamenshchik, C.Kiefer, A.A.Starobinsky and C.Steinwachs to whom
I am deeply grateful for a productive long term collaboration. I
also benefitted from helpful discussions with F. Bezrukov, G. Dvali,
J.~Barbon, J.~Garriga, C.~Germani, V.~Mukhanov, M.~Shaposhnikov and
S.~Solodukhin. I wish to express my gratitude for hospitality of the
Yukawa Institute for Theoretical Physics during the workshop
``Gravity and Cosmology" and the international symposium ``Cosmology
-- the Next Generation". This work was supported in part by the
Humboldt Foundation at the Ludwig-Maximilians University in Munich
and by the RFBR grant No 08-02-00725.

%

\end{document}